
%
%
\input psfig
\def\ce{\centerline}
\def\cite#1{[#1]}
\def\ref#1{(#1)}

\def\label#1{\eqno(#1)}
\font\largest=cmr10 scaled\magstep3
\font\larger=cmr10 scaled\magstep2

\vskip2in
{\largest
\centerline{On the transition from crystalline}
\vskip0.1in
\ce{ to }
\vskip0.1in
\ce{dynamically triangulated random surfaces}}
\vskip1in
{\larger\baselineskip15pt\ce{Neil Ferguson\footnote{$^1$}{{\rm e-mail:
ferguson@thphys.ox.ac.uk}} and
J.F. Wheater\footnote{$^2$}{{\rm e-mail: jfw@thphys.ox.ac.uk}}}
\ce{Department of Physics}
\ce{Theoretical Physics}
\ce{1 Keble Road}
\ce{Oxford OX1 3NP, UK
}}
\vskip1in
\noindent{\it Abstract}\break\noindent
We consider methods of interpolating between the crystalline and
dynamically triangulated random surface models.  We argue that actions
based on the deviation from six of the coordination number at a site
are inadequate and  propose an alternative based on Alexander moves. Two
simplified models, one of which has a phase transition and the
other of which does not, are discussed.
\vfil
\line{Theoretical Physics preprint  OUTP-93/17P\hfill 6th August 1993}
\eject

\def\bX{{\bf X}}

\def\bn{{\bf n}}

\def\expect#1{\langle {#1} \rangle}

\def\half{{1\over 2}}
\def\ij{<ij>}
\def\Rg2{R^2_g}

\def\Det{{\rm Det}}
\def\G{{\cal G}}
\def\JPA{1}
\def\kantor{2}
\def\ambj{3}
\def\jfwrgh{4}
\def\ukqcd{5}
\def\ambdurfro{6}
 \def\Fdavid{7}
 \def\Kazakov{8}
\def\polystring{9}
\def\ambjdurhuus{10}
\def\ADFJ{11}
\def\irback{12}
\def\bowick{13}
 \def\desmond{14}
\def\adfo{15}
 \def\boulatov{16}

Many apparently disparate physical systems,
 such as non-critical string theory, 2D quantum
gravity, and polymer membranes, are described by the statistical
 mechanics of random surfaces (for  recent reviews see \cite{\JPA}).
 This letter is concerned with understanding the
differences, if any, in the critical behaviour of crystalline surfaces
 and dynamically
triangulated surfaces.

Crystalline surfaces are described by a fixed
triangulation $T$ (in continuum terms a fixed intrinsic metric), with
the partition function  formed by summing over all the possible
embeddings of this surface in a $D$-dimensional euclidean space [\kantor];
 usually the fixed triangulation is a regular one but this is not
 essential.
If the lattice sites $i$ of $T$ have coordinates $\bX_i$ in
 the $D$-dimensional embedding space  the action is given by
$$ S_T(\kappa,D)=\half\sum_{\ij\in T}
 (\bX_i-\bX_j)^2+\kappa\sum_{\triangle\triangle'\in T}1-\hat\bn_\triangle.
\hat\bn_{\triangle'}\label{1}
$$
where $\ij$ denotes the link fron $i$ to $j$ and
 $\hat\bn_{\triangle,\triangle'}$ are the unit normal
vectors of the triangles on either side of the link.
The canonical partition function is
$$ Z_T(\kappa,D) =
\prod_{i\in T}\int d^D\bX_i\thinspace \delta^D\left(\sum_{j\in T}
 \bX_j\right) \thinspace e^{-S_T(\kappa,D)}
\label{2}
$$
where the delta function suppresses the translational
 zero mode. The first term in the action is the  elastic contribution
  while the second term which is proportional
   to the extrinsic curvature
of the surface inhibits bending. The model  exhibits the crumpling
transition, characterised by the Hausdorff dimension of the random
surface changing from $d_{H}=2$ (the smooth phase) to
$d_{H}\rightarrow\infty$ (the crumpled phase) at some critical value of
the extrinsic curvature coupling.  The existence of this transition,
which appears to be of second order,
is not proven analytically but is well
 established by the numerical studies [\kantor,\ambj,\jfwrgh,\ukqcd].

In the case of a dynamically
triangulated surface (DTRS), the canonical partition function
is formed by summing  over all possible embeddings {\it and}
all possible distinct
triangulations $T$ with a fixed number of points $N$ keeping one
  point marked
$$
Z_{N}\!(\kappa,D) =
\sum_{T:|T|=N}\rho(T) Z_T(\kappa,D)\label{3}
$$
where  $\rho(T)$ is a weighting factor depending
only  upon
the local properties of $T$ [\ambdurfro,\Fdavid,\Kazakov].
The grand canonical partition function  is then
$$
{\cal Z}(\mu ,\kappa,D) = \sum_{N=1}^{\infty } e^{-\mu\! N } Z_{N}
\! (\kappa,D)\label{4}
$$
 The sum over
triangulations is equivalent to the integral over worldsheet metrics in
Polyakov's (\cite{\polystring}) formulation of string theory; hence
 these models
are used in the numerical study of non-critical strings and 2-D quantum
gravity.
 It is known that in order
  to obtain  an interesting continuum
 limit in which  both the string tension and
  the mass gap  scale to zero it is necessary (but not
   necessarily sufficient) to have  the extrinsic curvature term in the
   action [\ambjdurhuus,\ADFJ].
    In the last few years much computational effort has been
expended on these models but
	the results  of these simulations  are  much less clear-cut
 than for the
 crystalline surface.  The  Monte-Carlo simulation
  of systems with dynamical triangulation  is very
computationally intensive which limits the size of
lattice that can be studied and the difficulties are
 compounded by the fact that the crumpling transition,
  if it exists,  seems to be rather weak with only a cusp
   singularity in the specific heat \cite{\irback,\bowick}.

As an alternative to studying the DTRS directly we consider
 a generalized model which interpolates between  it and
  the better understood crystalline surface.
 Introduce the notion of a distance, $n(T_0,T)$, between
   a general triangulation $T$ and  reference triangulation $T_0$
   (we shall consider shortly what form this distance might take)
 and consider the  canonical partition function for
a generalized model
$$
Z_N'(\zeta,\kappa,D) =\sum_{T:|T|=N}{\rho(T)}
 \exp\left(-\zeta n(T_0,T) \right) Z_T(\kappa,D)\label{5}
$$
When $\zeta = 0$ this coincides with the DTRS
 while as $\zeta\to\infty$ all triangulations except
  $T_0$ are suppressed and we obtain the canonical
   partition function for the crystalline surface  on $T_0$.
    It is slightly more complicated to define a
grand canonical partition function for this new model
 because not all forms of
reference triangulation $T_0$ can be written naturally for all $N$.
   However, in the cases where a natural definition is possible, such
as the example we will discuss later,
  the grand canonical partition function is given by
$$
{\cal Z}'(\mu ,\zeta,\kappa,D) = \sum_{N=1}^{\infty } e^{-\mu\! N }
 Z'_{N}
\! (\zeta,\kappa,D)\label{6}
$$

At large $N$ the asymptotic behaviour of $Z_N$ is expected to be
$$Z_N=N^{\gamma-2} e^{\mu_c N}\left(1+O\left({1\over N}\right)\right)
\label{7}$$
Accordingly the grand canonical partition function develops
 non-analytical behaviour as $\mu\downarrow \mu_c$ and the
 critical exponent $\gamma$ controls
the derivative of $\cal Z$ in this limit through
$${\partial {\cal Z}\over \partial\mu} \sim {1\over
(\mu-\mu_c)^\gamma}\label{8}
$$

The phase
 diagram of the generalized model in the $(\kappa,\zeta)$ plane
 certainly
 depends on $D$; in practice, most work on the crystalline surface
 and the
DTRS has been done in $D=3$.
It is known that the crystalline surface has a crumpling transition
 at
$\kappa=\kappa_c$; for $\kappa<\kappa_c$ a surface of $N$ sites
 has mean
square radius $R^2\sim \log N$ whereas for $\kappa>\kappa_c$,
 $R^2\sim N$.
Since there is no smooth interpolating function between $N$ and
$\log N$
when $N\to\infty$ these two regions must be separated by  a
 phase
transition. The simplest possibility is that there is a line
 of
transitions (the solid line in fig.1) extending all the way
 across the
diagram and cutting the $\zeta=0$ axis at $\kappa_c'$; in this
 case the
DTRS also has a crumpling transition.  However there may be a
critical point at
X so that the order of the transition is different  for the
 DTRS; there is no
numerical evidence for a first order transition but a higher
order one is certainly possible. Alternatively,
the line may stop at an
intermediate point X provided that one or both of the lines of
 transitions
AB, CD is present to isolate the two phases of the crystalline
surface
(the order of A, C in the  picture is of no significance and
 they might coincide with X); in this case
there is no crumpling transition in the DTRS but there may be
some
non-trivial behaviour arising from the proximity of X to the
 $\zeta=0$ line
(it was pointed out in [\bowick] that the
 observed behaviour is quite similar to
that seen in models which are known not to have a phase
 transition but
rather display some crossover behaviour). Furthermore,
 since it is
certainly known that $R^2\sim N^\nu$ with $\nu \approx 0$
when
$(\kappa,\zeta)=(0,0)$, if the crumpling transition line
 does stop at X
then the line of transitions CD must exist.  Assuming that
 actually
$\nu=0$, the existence of the line AB cannot be deduced from
 the mean
square radius; however there are other quantities, such as
 $\gamma$,
which can differentiate between phases. The rest of this
 paper is
concerned with the behaviour along the $\kappa=0$ line.

A recent paper [\desmond]  proposed using
 the coordination number of the vertices
as a measure of the distance $n(T,T_0)$.  In a regular
triangulation all
(or almost all, depending upon the genus) the vertices
 have coordination
number $q_i=6$; thus we could choose
$$
n(T,T_0)=\sum_{i\in T} \vert q_i-6 \vert^\alpha\label{9}
$$
for some positive power $\alpha$ to give the distance of $T$
 from a regular
triangulation. In [\desmond] it was argued that $\alpha\ge2$
corresponds to an
irrelevant operator in the continuum limit and so  $\alpha=1$
was chosen.  However, it is clear that for $\alpha\ge2$,  $n(T,T_0)$
increases faster as $T$ becomes more different from $T_0$ than
 it does for
 $\alpha=1$; thus it is hard to see how  $\alpha=1$ can be
{\it more}
 effective at suppressing the irregular triangulations
regardless of
 considerations about the continuum limit. Let us assume
 that the suppression
of irregular triangulations is strong enough that when
 $\zeta\to\infty$ only
the most regular possible triangulations contribute.
 For genus zero there are many
 triangulations with $\{q_i=6,\thinspace \forall i\}$ differing
 only by
  their modular
 parameter which can vary from $\tau=1$ ({\it ie} the shortest
 cycle is
 $O({\sqrt N})$ links long) to $\tau=O(N)$ ({\it ie} the
 shortest cycle is 3
links long);
  they all contribute to $Z_N$
 and are not suppressed at all by the $\zeta n(T,T_0)$ term
 in the action.
 Thus even at $D=0$ the action \ref{9} is not suitable to
 interpolate
 between the fluid and the crystalline surfaces. At non-zero
 $D$ the $\bX$
 integral in \ref{2} can be done when $\kappa=0$ to get
  [\ambdurfro,\Fdavid,\Kazakov]
 $$
 Z_N(0,D)=\sum_{T:|T|=N}\rho(T) \left( \Det' I_T\right)^
{-{D\over 2}}
 \label{10}
 $$
 where $I_T$ is the incidence matrix of $T$.  At large $N$
 we know that
 $$
 \left( \Det' I_{T(\tau=O(N))}\right)^{-1}\sim e^{\beta N}
\left(
  \Det' I_{T(\tau=1)}\right)^{-1}
   \label{11}
  $$
 for finite positive $\beta$ [\adfo].  This means that in the
 thermodynamic limit
 the ultra-thin configurations will dominate in \ref{10}.
 Hence, even
 in the limit $\zeta\to 0$, the partition function will be
 dominated not
 by a nice regular two-dimensional surface but rather by quasi
 one-dimensional spaghetti-like objects.  We believe that this
 phenomenon
 may be responsible, in part, for the observation in [\desmond]
 that, although
 the intrinsic geometry is smoothed by \ref{9}, the extrinsic
 geometry is not.
 Clearly the long thin tubes will be suppressed by the extrinsic
 curvature term when $\kappa >0$. However, the extrinsic
curvature can certainly not be relied upon to suppress all the
triangulations except one; in addition the $\kappa$ region where
 the spaghetti freezes out could get mixed up with the crumpling
 transition of the $\tau\approx 1$ triangulations and add
 to the confusion.

 To make progress a better definition of $n(T,T_0)$ is needed.
It is known that
any two triangulations  $T_0,\thinspace T$ with the same
 number of vertices and the same topology are related
 by a sequence of link flip operations (known as Alexander moves).
Let us define $n(T,T_0)$ to be {\it the minimum number of flips
 necessary
to change $T$ into $T_0$}. (It is doubtful that the absolute
minimum is
really needed; the number of flips taken by any well-defined
reproducible
procedure will do as well.) The new definition of $n(T,T_0)$
will get rid of the
spaghetti. It takes $O(N)$ flips to turn a $\tau=1$ triangulation
into one
with $\tau=O(N)$ so the suppression factor $e^{-\zeta n}$ can
always be made to
overcome the effect of \ref{11} by choosing $\zeta$ large enough.
This definition of $n(T,T_0)$ seems much harder to use than
 \ref{9}, especially in
a numerical simulation; however there are some special cases
(unfortunately
in $D=0$)  which can be solved and we now turn to them.

It is convenient to work in terms of the dual graphs rather
 than the
triangulations themselves. We first consider three point
functions and take as
our reference graph $G_0(N)$ the one shown in fig.2  (note
that the actual number of points in the graph is $2N+1$).
 It was demonstrated in
[\boulatov]
that any three point function graph can be reduced to this
 form by
flips.  Of course such a reference graph is very different
from the one on which
we would like to define the crystalline surface but it serves
 to illustrate the principle. In the sum over the graphs we
 include all those which can be
constructed from tree graphs in the manner shown in fig.3;
 ${\cal T}_k$ denotes
the set of rooted trees with $k$ branches and it is known
[\boulatov] that the
number of graphs, $T_k$, in this set is given by the $k$'th
  Catalan number
$$
T_k={(2k-2)!\over k! (k-1)!}
   \label{12}
  $$
It is easy to see that the rooted tree ${\cal T}_k$ in a graph
 of the form shown in fig.3 can be reduced to the
form of the reference graph by making $k-1$ flips, each time
flipping the
link that is the trunk of a tree. Now introduce the generating
 function
$$
F_N(t)=\sum_{n=0}^{n_{max}} \Omega_{N,n} t^n
   \label{13}
  $$
 where $\Omega_{N,n}$ is the number of graphs which can be moved to the
 reference graph $G_0(N)$ by making $n$ flips and $t\equiv e^{-\zeta}$.
 The graphs obey the Schwinger-Dyson equation shown in fig.4 and hence
 $F_N(t)$ obeys the recursion relation
$$
F_N(t)=\sum_{k=1}^{N}  t^{k-1} T_k F_{N-k}(t)
   \label{14}
  $$
 Multiplying by $z^N$ ($z\equiv e^{-\mu}$) and summing over $N$ gives,
 after some manipulation,
$$
\G(z,t)=\sum_{N=0}^\infty F_N(t)=1+t^{-1}\G(z,t)\sum_{k=1}^{\infty}  (zt)^k
T_k
   \label{15}
$$
The last sum is just the generating function for the Catalan numbers so we
find that the grand canonical partition function $\G(z,t)$ is given by
$$
\G(z,t)={2t-1-\sqrt{1-4zt}\over 2(z-1+t)}
   \label{16}
$$
$\G(z,t)$ displays non-analyticity as we take  $z\to z_c(t)$ to obtain the
thermodynamic limit; we find the following behaviour
$$
\eqalign{
t&<\half,\quad z_c=1-t\quad \G\sim (z_c-z)^{-1}\cr
t&=\half,\quad z_c=\half\qquad\;\, \G\sim
(z_c-z)^{-\half}\cr
t&>\half,\quad z_c={1\over 4t}\qquad{\partial \G\over\partial z}
\sim (z_c-z)^{-\half}
}
   \label{17}
 $$
 The behaviour for $t>\half$ corresponds to
 $\gamma=\half$ which is the value expected for a system of
 branched polymer-like  configurations; given the method of construction
 of our set of graphs that is not surprising.  For $t<\half$, $\G$
 itself diverges; at $t=0$ it
 is easy to check that we get exactly the partition function expected
 when only the reference graphs $G_0(N)$ are included demonstrating that
 varying $t$ does indeed allow us to interpolate between an ensemble of
 graphs and a fixed reference graph.  The standard thermodynamic
 quantities are easy to calculate and we find
 $$
 {1\over N}\expect{n(G_,G_0)}=\cases{ {t\over 1-t}&$t\le\half$\cr
1&$t>\half$\cr
}
 \label{18}
$$
and

 %
 $$
 {1\over N}\expect{n(G_,G_0)^2}_C =\cases{ {t\over (1-t)^2}&$t\le\half$\cr
 0&$t>\half$\cr
}
 \label{19}
$$
There is no divergent specific heat or other behaviour to suggest that a
correlation length is diverging at the critical point $t=\half$.

  The model
 can be generalized slightly to include all graphs made of
  ${\cal T}_k$ as shown in fig.5 (considering the four point function in
this case simplifies the calculation).  By making a succession of flips
  as indicated this can be turned into
a graph in which the root of ${\cal T}_k$ is isolated as before.  The number
of flips to do this is $2\min(q-1,k-q)$ so that the factor $T_k$ in \ref{14}
is replaced by
$$T_k\times\sum_{q=1}^k t^{2\min(q-1,k-q)}\label{20}$$
(In fact this does not give us the {\it minimum} number of flips to
convert these graphs to $G_0$ form but it does yield a soluble model.)
  Similar manipulations as before lead to the result
$$
\G(z,t)={t^2(1-t^2)\over t^2(1-t^2)-t\left(\sqrt{1-4zt^2}-\sqrt{1-4zt}\right)
+{1\over 4}(1-t)^2\left(\sqrt{1+4zt^2}-\sqrt{1-4zt^2}\right)}\label{21}
$$
Despite its apparent complexity the non-analytic behaviour of $\G$ is
 the same for all $t$ and takes the form
$$\G(z,t)\sim {1\over z_c-z}\label{22}$$
There is no phase transition; there is no need for one because
 $\gamma$, the only order parameter that could distinguish between
  $t=0$ and $t=1$ actually
takes the same value in each case.

Although our definition of $n(T,T_0)$ has some appealing properties it is
 hard to see how it might be implemented efficiently in a Monte Carlo
 simulation to study the evolution of the crystalline surface into the
 dynamically triangulated one.  To take a random triangulation and ``flatten"
 it into a regular $T_0$ by means of flips in the minimum possible number
 of moves seems to be a problem
of the travelling salesman type.  It is easy enough to count flips going away
from $T_0$  so long as their concentration is small but this is hardly
 likely to
be the interesting regime for the transition between the two surface models.

\vskip0.5in\noindent
We are grateful to Thordur Jonsson for many valuable conversations on
this subject. N.F. ackowledges an SERC graduate studentship and this work
was supported in part by SERC grant GR/H01243.
\vfill

\line{\larger References}
\def\nl{\hfil\break}
\def\refer#1#2{\item{[#1]}{#2}}
\smallskip
\refer{\JPA}{``Simplicial  quantum gravity and random lattices", F.David,
Saclay Physique Theorique preprint 93/028;\nl
``Random surfaces: from polymer membranes to strings", J.F.Wheater,
Oxford Theoretical Physics preprint OUTP-93/11P.}
\refer{\kantor}{Y.Kantor and D.Nelson, Phys. Rev. Lett. 58 (1987) 2774;\nl
Phys. Rev. A 36 (1987) 4020.}
\refer{\ambj}{J.Ambj\o rn, B.Durhuus and T.Jonsson,
  Nucl. Phys. B 316 (1989) 526.}
\refer{\jfwrgh}{R.G.Harnish and J.F.Wheater, Nucl. Phys. B 350 (1991) 861.}
%

\refer{\ukqcd}{J.F.Wheater and P.W.Stephenson, Phys. Lett. B302 (1993) 447. }

\refer{\ambdurfro}{J.Ambj\o rn, B.Durhuus and J.Fr\"ohlich,
Nucl. Phys. B 257 [FS14](1985) 433.}
\refer{\Fdavid}{F.David,
Nucl. Phys. B 257 [FS14](1985) 543.}
\refer{\Kazakov}{V.A.Kazakov, I.K.Kostov and A.A.Migdal, Phys. Lett. 157B
(1985) 295.}
\refer{\polystring}{A.M.Polyakov, Phys. Lett. 103B (1981) 207.}
\refer{\ambjdurhuus}{J.Ambj\o rn and  B.Durhuus, Phys. Lett. B188
 (1987) 253.}
\refer{\ADFJ}{J.Ambj\o rn, B.Durhuus, J.Fr\"ohlich and T.Jonsson,
Nucl. Phys. B 290 [FS20](1987) 480.}
\refer{\irback}{J.Ambj\o rn, J.Jurkiewicz, S.Varsted and A.Irb\"ack,
 Phys. Lett.  275B (1992) 295;\nl
J.Ambj\o rn, A.Irb\"ack, J.Jurkiewicz and  B.Petersson, Nucl. Phys.
 B393 (1993) 571.}

\refer{\bowick}{M.Bowick {\it et al}, Nucl. Phys. B394 (1993) 791.}

\refer{\desmond}{
``Freezing a fluid random surface", C.F.Baillie and D.A.Johnston,
 preprint COLO-HEP-316,\nl hep-lat/9305012.}
\refer{\adfo}{J.Ambj\o rn, B.Durhuus, J.Fr\"ohlich and P.Orland,
Nucl. Phys. B 270 [FS16](1986) 457.}
\refer{\boulatov}{D.V Boulatov, V.A Kazakov, I.K.Kostov
 and A.A Migdal, Nucl. Phys. B275[FS17] (1986) 641}

\vskip1in
\line{\larger Figure Captions\hfil}
\medskip
\item{1)}{Illustrating possible phase diagrams for the generalized
 model.}
\item{2)}{The reference dual graph $G_0(N)$.}
\item{3)}{Restricted set  of three point graphs constructed from
tree graphs.}
\item{4)}{Schwinger-Dyson equation for the restricted set of three
 point graphs.}
\item{5)}{A larger set of graphs constructed from tree graphs.}

\vfill\eject
\def\ce{\centerline}

\ce{\vbox to 3.5in{\hbox to 5.35in{
\psfig{figure=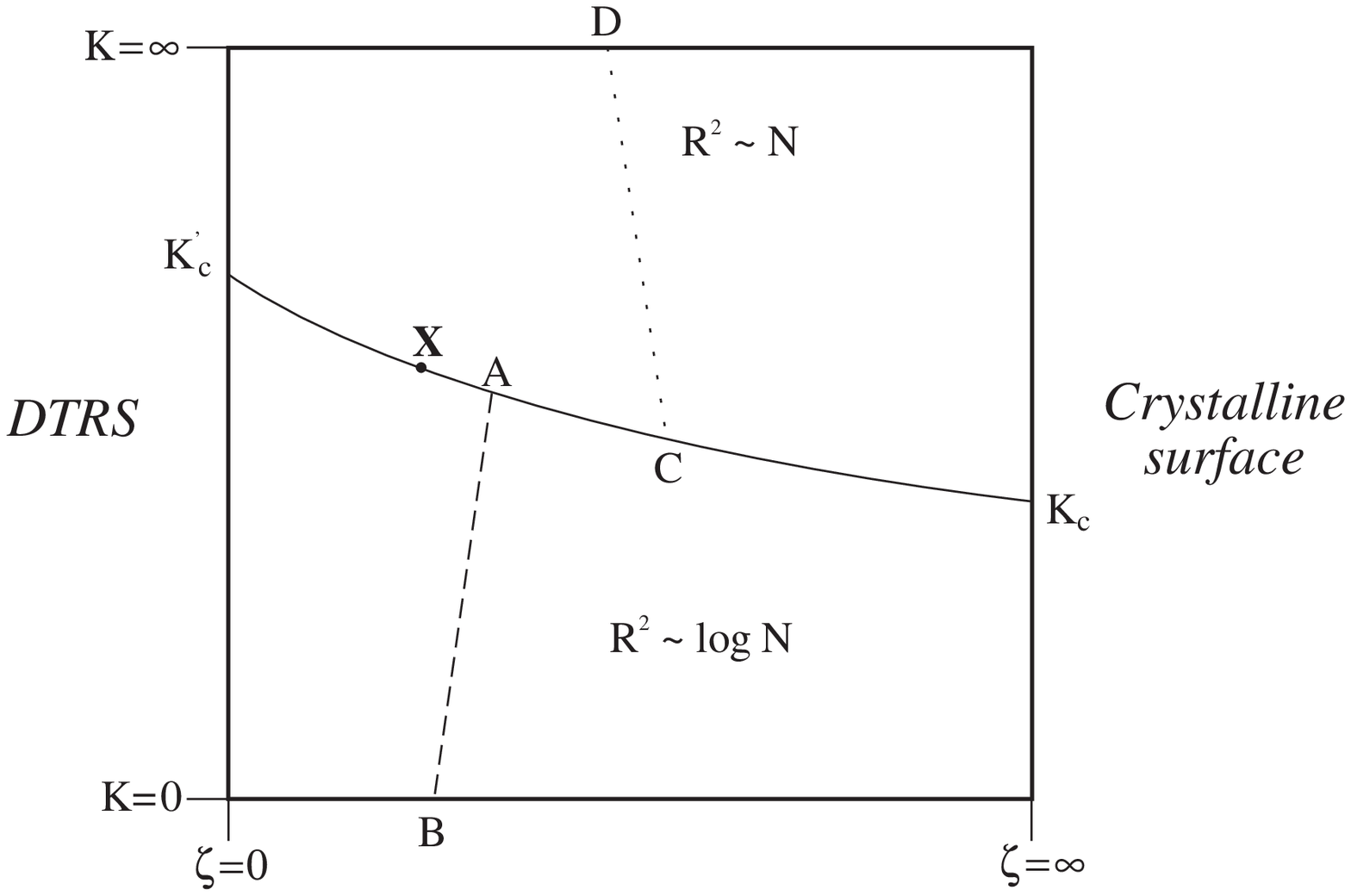,height=3.5in}}}}
\ce{fig.1}
\vskip0.7in

\ce{\vbox to 2.5in{\hbox to 2.45in{
\psfig{figure=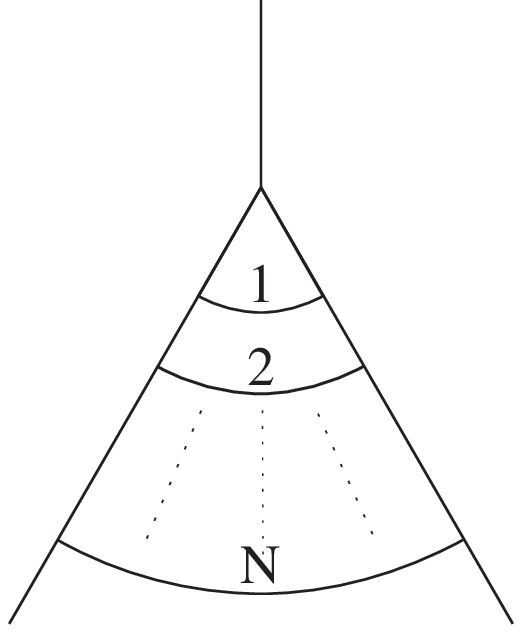,height=2.5in}}}}
\vskip0.2in
\ce{fig.2}
\vskip0.5in

\ce{\vbox to 2.5in{\hbox to 4.55in{
\psfig{figure=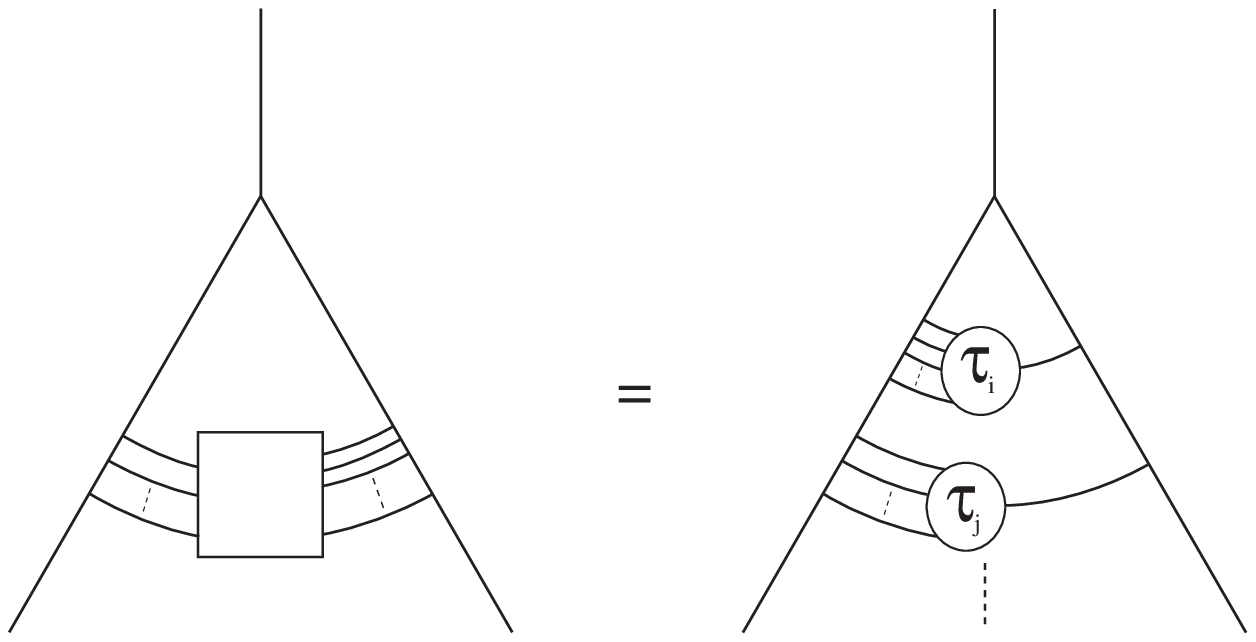,height=2.5in}}}}
\vskip0.2in
\ce{fig.3}
\vskip0.5in

\ce{\vbox to 2.5in{\hbox to 5.35in{
\psfig{figure=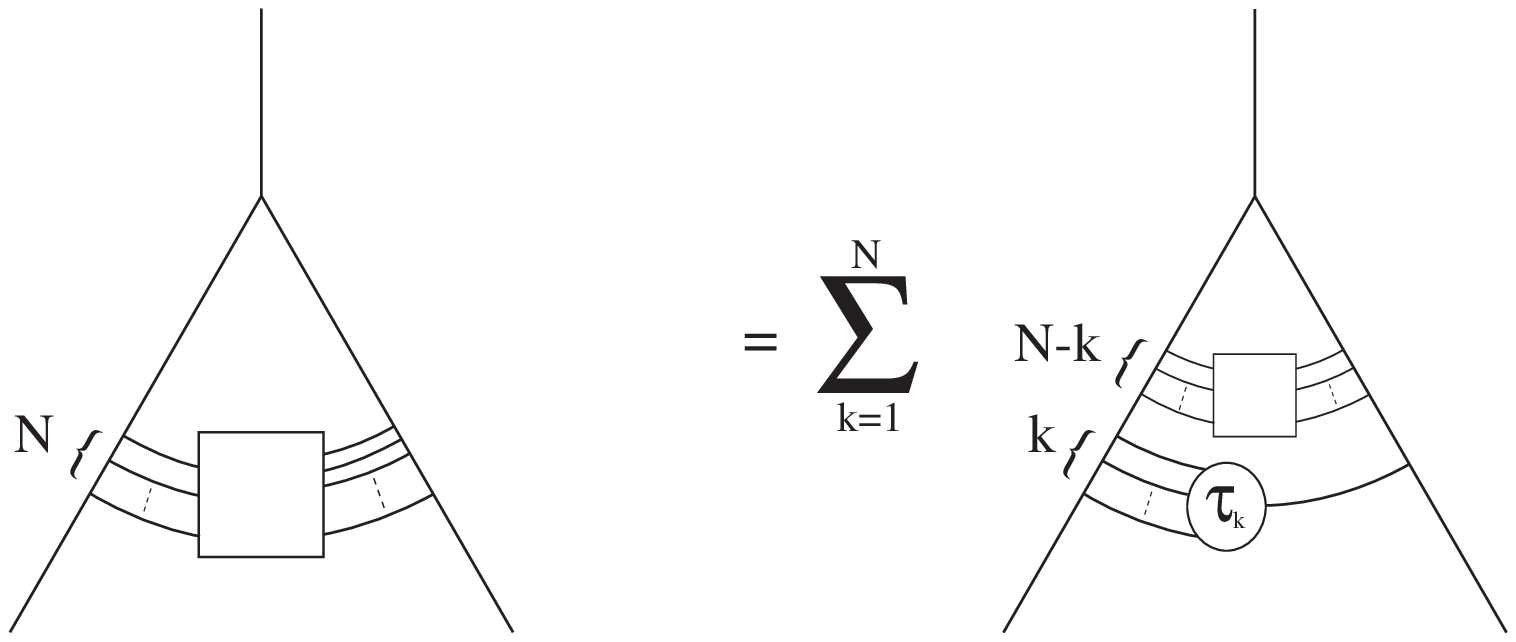,height=2.5in}}}}
\vskip0.2in
\ce{fig.4}
\vskip0.5in

\ce{\vbox to 1.7in{\hbox to 5.95in{
\psfig{figure=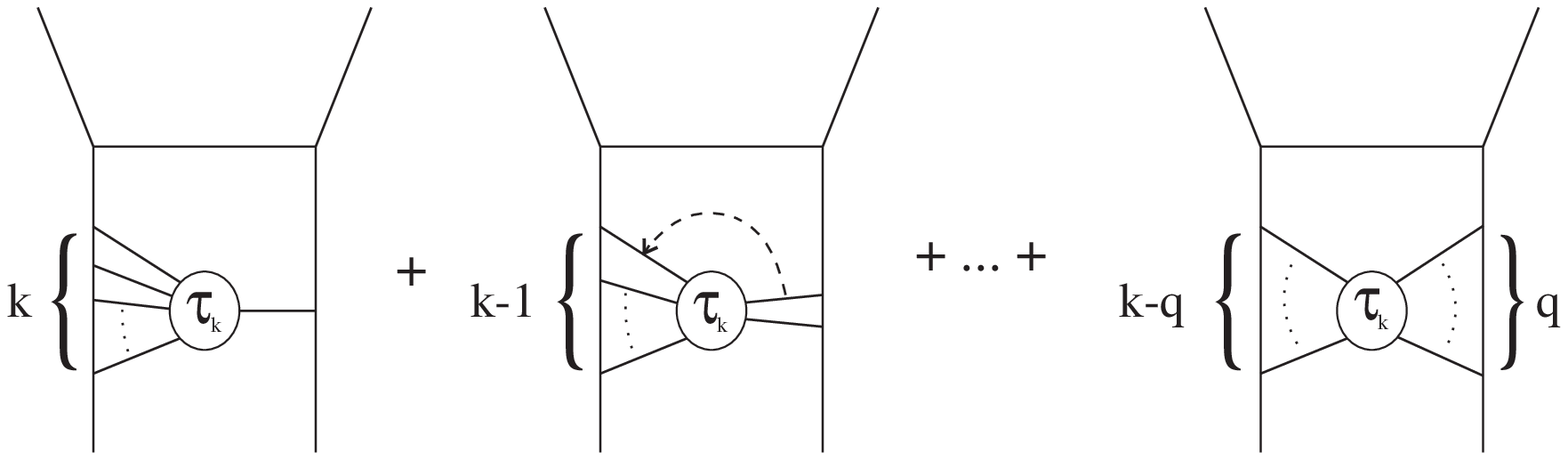,height=1.7in}}}}
\vskip0.2in
\ce{fig.5}
\vskip0.5in

\bye